\pdfoutput=1

\documentclass{new_tlp}
\usepackage{mathptmx}
\usepackage{afterpage}

\usepackage{xspace}
\usepackage{alltt}
\usepackage[T1]{fontenc}
\usepackage{url}
\usepackage{tablefootnote}

\newcommand\mycode[1]{\texttt{\small #1}} 
\newcommand\defn[1]{\emph{#1}} 
\newcommand{\footnoteref}[1]{\textsuperscript{\ref{#1}}}

\newcommand\m[1]{\mbox{$#1$}} 
\newcommand\IF{\m{\leftarrow}\xspace}
\newcommand\THEN{\m{\rightarrow}\xspace}
\newcommand\bigO[1]{\m{O(#1)}} 
\newcommand\sinO[1]{\m{O(#1)}} 
\newcommand\sincO[1]{\sinO{\mycode{#1}}}  
\newcommand\sincsO[1]{\sincO{\##1}} 
\newcommand\doubcsO[2]{\sincO{\##1$\times$\##2}} 
\newcommand\ofour[4]{\m{O(\mycode{min}(\mycode{\##1}\times\mycode{\##2},\ \mycode{\##3}\times\mycode{\##4}))}}
\newcommand\vpt{v\_pt}
\newcommand\fpt{f\_pt}
\newcommand\apt{a\_pt}
\newcommand\tpt{t\_pt}
\newcommand{\myparag}[1]{\paragraph{\rm \bf #1.}}
\newenvironment{code}{\begin{alltt}\footnotesize}{\end{alltt}}

\newcommand{\notes}[1]{} 

\newcommand{\tplpadd}[1]{}
\newcommand{\tplpdel}[1]{#1}

\newcommand\bcmdtab{\noindent\bgroup\tabcolsep=0pt%
  \begin{tabular}{@{}p{10pc}@{}p{20pc}@{}}}
\newcommand\ecmdtab{\end{tabular}\egroup}

\jdate{ICLP 2016}
\pagerange{35--48}
\pubyear{2016}
\volume{\textbf{8} (1):}
\pagerange{\pageref{firstpage}--\pageref{lastpage}}
\doi{---}
  \title[Precise Complexity Guarantees for Pointer Analysis via Datalog with Extensions]
        {Precise Complexity Guarantees for Pointer Analysis via Datalog with Extensions\thanks{
    This work was supported in part by NSF under grants
    CCF-1414078, 
    IIS-1447549, 
    CCF-1248184, 
    CCF-0964196, 
    and ONR under grant
    N000141512208. 
        }}

  \author[K. T. Tekle, Y. A. Liu]
         {K. TUNCAY TEKLE, YANHONG A. LIU\\
         Computer Science Department, Stony Brook University, Stony Brook, NY, USA\\
         \email{tuncay,liu@cs.stonybrook.edu}}

\begin{document}

\label{firstpage}

\maketitle

  \begin{abstract}
Pointer analysis is a fundamental static program analysis for computing the set of objects that an expression can refer to.  Decades of research has gone into developing methods of varying precision and efficiency for pointer analysis for programs that use different language features, but determining precisely how efficient a particular method is has been a challenge in itself.

For programs that use different language features, we consider methods for pointer analysis using Datalog and extensions to Datalog. When the rules are in Datalog, we present the calculation of precise time complexities from the rules using a new algorithm for decomposing rules for obtaining the best complexities. When extensions such as function symbols and universal quantification are used, we describe algorithms for efficiently implementing the extensions and the complexities of the algorithms.
  \end{abstract}


\section{Introduction}
\label{sec:intro}

Pointer analysis is a static program analysis for computing the set of objects that an expression can refer to. It is a fundamental analysis used for many applications, e.g., debugging~\cite{DBLP:conf/sas/ShapiroH97}, performance analysis~\cite{DBLP:conf/pldi/GhiyaLS01}, dataflow analysis~\cite{DBLP:conf/sas/ShapiroH97}, parallelism~\cite{DBLP:conf/pldi/WilsonL95,DBLP:journals/toplas/PearceKH07}, common subexpression elimination~\cite{DBLP:conf/pldi/DiwanMM98,DBLP:conf/popl/GhiyaH98}, optimization by incrementalization~\cite{Gor+10Alias-DLS}, and detection of security vulnerabilities~\cite{DBLP:conf/icse/AvotsDLL05}.
Consider the following program fragment in an object-oriented programming language:

\begin{code}
void foo() \{
  Object o1 = new Object();
  Object o2;
  if (...) o2 = id(o1);
  else o2 = new Object();
\}
void id(Object o) \{ return o; \}
\end{code}

For each variable \mycode{o1}, \mycode{o2}, and \mycode{o}, a pointer analysis may aim to find the set of objects the variable \emph{may} point to. We call each such set a \emph{may-point-to} set. A may-point-to analysis is \emph{sound} if each may-point-to set contains all objects that the variable may point to at runtime. A sound may-point-to analysis is \emph{precise} if no may-point-to set contains more objects than the set of objects that the variable may point to at runtime. Another form of analysis is \defn{must-point-to} analysis, which finds the set of objects that each variable \emph{must} point to. Precise pointer analysis is undecidable~\cite{DBLP:journals/loplas/Landi92,DBLP:journals/toplas/Ramalingam94}. Therefore, pointer analysis methods approximate the results, providing different tradeoffs between the precision of the results and the efficiency of the method~\cite{DBLP:conf/issta/HindP00} while preserving soundness. Whereas may-point-to analysis is an overapproximation, must-point-to analysis is an underapproximation.

From the simple program fragment above, it can be seen that the precision-efficiency tradeoff involves the consideration of control flows, procedures, calling contexts, objects, and other language features such as arrays; leading to a variety of analyses. The worst-case time complexities of existing analyses vary from almost linear~\cite{DBLP:conf/popl/Steensgaard96} to doubly exponential~\cite{DBLP:journals/toplas/SagivRW98}. However, such worst-case complexities are often not a true indication of analysis time; many researchers provide empirical performance results for their algorithms, and many papers have been written on which pointer analysis one should use, e.g.,~\cite{DBLP:conf/issta/HindP00}, including one questioning whether we have solved the pointer analysis problem yet~\cite{DBLP:conf/paste/Hind01}.

A recent survey~\cite{DBLP:journals/ftpl/SmaragdakisB15} presents existing work on logical specifications of pointer analysis methods in the declarative language \emph{Datalog} and its extensions.
Datalog specifications allow expressing the precision aspects of pointer analyses concisely, while ensuring that the analyses are performed in polynomial time, because evaluation of rules in Datalog is guaranteed to be polynomial time. However, just as the worst-case complexities of existing analyses are not a true indication of analysis time, worst-case polynomial time for evaluation of Datalog is not sufficient for understanding actual running times.

In this paper, we consider all the different analyses presented in the survey~\cite{DBLP:journals/ftpl/SmaragdakisB15} expressed in Datalog and its extensions, and study the time complexity of each analysis by using and extending a systematic method for calculating the time complexities for optimal bottom-up evaluation of Datalog rules~\cite{Liu:2009:DRE:1552309.1552311}. To obtain the best complexity, we give a new algorithm for rule decomposition. All analyses require handling rules with many hypotheses, and some require handling extension to Datalog with negation, function symbols, and universal quantification. In each case, we describe the method for handling the extensions and calculating the time complexities. Our methods can be readily used for analyzing other program analyses expressible in Datalog and similar extensions.


\section{Language and preliminaries}
\label{sec:prelim}

In this section, we describe Datalog, an optimal method for evaluating a set of Datalog rules with at most two hypotheses each, and a method for calculating the time complexity of the evaluation.

\myparag{Datalog}

\defn{Datalog} is a language for defining rules, facts, and queries, where rules can be used with facts to answer queries.  A Datalog rule is of the form:
\[
p(a_1,\textrm{...},a_k)~ \IF~ p_1(a_{11},\textrm{...},a_{1k_1}),\textrm{...},\  p_h(a_{h1},\textrm{...},a_{hk_h}).
\]
where $h$ is a finite natural number, each $p_i$ (respectively $p$) is a predicate of finite number $k_i$ (respectively $k$) arguments, each $a_{ij}$ and $a_i$ is either a constant or a variable, and each variable in the arguments of $p$ must also be in the arguments of some $p_i$.
If $h=0$, then each $a_i$ must be a constant, in which case $p(a_1,\textrm{...},a_k)$ is called a \defn{fact}.
For the rest of the paper, ``rule'' refers only to the case where $h\geq 1$, in which case each $p_i(a_{i1},\textrm{...},a_{ik_i})$
is called a \defn{hypothesis}, and $p(a_1,\textrm{...},a_k)$ is called the \defn{conclusion}.
For rules with the same hypotheses but different conclusions, we use the shorthand of writing one rule with the same hypotheses but with comma-separated conclusions.

The meaning of a set of rules and facts is the set of facts that are given or can be inferred using the rules.

\myparag{Terminology}

An \defn{IDB (intensional database) predicate} is a predicate defined by rules, and an \defn{EDB (extensional database) predicate} is a predicate for which no rules exist, and only facts are given. An \defn{IDB (EDB) hypothesis} is a hypothesis whose predicate is an IDB (EDB) predicate.

For complexity calculation, we use the following notations.
\begin{itemize}

\item \mycode{\#p}: number of facts of predicate \mycode{p}, called \defn{size of} \mycode{p}.

\item \mycode{\#dom(p.i)}: size of the domain from which the \mycode{i}th argument of predicate \mycode{p} takes value.

\item \mycode{\#p.i}: number of values actually taken by the $i$th argument of the facts of predicate \mycode{p} (given or inferred).

\item \mycode{\#p.i$_1$,\textrm{...},i$_n$/j$_1$,\textrm{...},j$_m$}: maximum number of combinations of different values actually taken by the \mycode{i$_1$},\textrm{...},\mycode{i$_n$}th arguments of the facts of predicate \mycode{p} (given or inferred), given any fixed value for the \mycode{j$_1$},\textrm{...},\mycode{j$_m$}th arguments.

\end{itemize}

We assume that hash tables, tries or similar data structures are used so that operations involving a single element of a set take \bigO{1} time.

\subsection{Bottom-up evaluation and complexity calculation}
\label{subsec:complexity}

Bottom-up evaluation starts with given facts, infers new facts from conclusions of rules whose hypotheses match existing facts, and does so repeatedly until all facts are inferred.  In this paper, we use the bottom-up evaluation method of~\cite{Liu:2009:DRE:1552309.1552311}. The time complexity incurred by each rule using this method is bound by the \defn{number of firings} of the rule---the number of combinations of facts that make all hypotheses true.  The best complexity is the minimum among all possible decompositions of the rule into rules with at most two hypotheses. However, the number of decompositions of a rule is worse than exponential in the number of hypotheses.

In this subsection, we summarize how to compute the optimal time complexity incurred by a rule with at most two hypotheses with the method in~\cite{Liu:2009:DRE:1552309.1552311}. Note that the number of rules and the arities of predicates are considered constants, not affecting the asymptotic analysis. In the next section, we describe a heuristic algorithm for decomposing rules so that each rule has at most two hypotheses and then calculating the optimal complexity for the decomposition.

There are two forms of rules when the rules are limited to two hypotheses. When a rule has one hypothesis, it is of the form: \mycode{p(...) \IF q(...).} The number of times this rule can fire is the number of facts of \mycode{q}, therefore the time complexity incurred by this rule is \sincsO{q}. In fact, we can omit the complexity of such rules, because (i) if \mycode{q} is an EDB predicate, then all of its facts need to be read in, therefore \sincsO{q} cost is already incurred by the reading of the input, (ii) if \mycode{q} is an IDB predicate, then its size would be bound by the complexity of the rules that infer its facts, and therefore that complexity would already have been included by the complexity of the rules inferring its facts.

When a rule has two hypotheses, it is of the form: \mycode{p(...) \IF q(x$_1$,...,x$_n$), r(y$_1$,...,y$_m$).} To calculate the number of firings, we can first think of processing the facts of \mycode{q} and matching them with facts of the second hypothesis such that the common variables in the hypotheses take the same value. Therefore, only the variables of the second hypothesis not in the first can take values for each fact of \mycode{q}. We use $C_{12}$ to denote the set of integers $j$ in $[1,..,m]$ such that \mycode{y$_j$} is a variable common to both hypotheses, then the complexity is bounded by \doubcsO{q}{r.$i \notin C_{12}/j \in C_{12}$}. Analogously, we can think of processing the facts of \mycode{r} and matching them with facts of the first hypothesis. Defining $C_{21}$ analogously to be the set of integers $j$ in $[1,..,n]$ such that \mycode{x$_j$} is a common variable, the complexity is also bounded by \doubcsO{r}{q.$i \notin C_{21}/j \in C_{21}$}. Since both bounds are upper bounds on the number of firings, then the complexity is bounded by the minimum of the two: \ofour{q}{r.$i \notin C_{12}/j \in C_{12}$}{r}{q.$i \notin C_{21}/j \in C_{21}$}. For example, for the rule \mycode{p(x,z) \IF q(x,y), r(y,z).}, the complexity is \ofour{q}{r.2/1}{r}{q.1/2}.

\section{Handling many hypotheses---applied to pointer analyses for different language features}
\label{sec:base}

In general, rules may have many hypotheses. In this section, we describe an algorithm for decomposing rules so that the resulting set of rules has at most two hypotheses and achieves the best complexity among all possible decompositions. We then apply our algorithm coupled with the complexity calculation of bottom-up evaluation to calculate the complexity of may-point-to analysis for object-oriented languages. The algorithm applies also to analyses of other advanced language features, including procedures, arrays, and exceptions as shown in Appendix A\tplpadd{ of~\cite{2016arXiv160801594T}}.

\subsection{An algorithm for decomposing rules with many hypotheses}
\label{subsec:decompose}

Given a set of rules where some rules have more than two hypotheses, each such rule can be \defn{decomposed} so that a set of rules with the same meaning is produced where each rule has two hypotheses. To decompose a rule $R$, we (1) select two hypotheses $h_1$ and $h_2$ of $R$; (2) create a new intermediate rule $R'$ whose hypotheses are $h_1$ and $h_2$, and whose conclusion $c$ is a new, intermediate predicate whose arguments are the variables occurring in $h_1$ or $h_2$ that are also elsewhere in $R$; (3) replace $h_1$ and $h_2$ in $R$ with $c$; and (4) repeat steps (1)--(3) until $R$ has only two hypotheses.

How two hypotheses are selected at step (1) of each iteration so that the running time of the resulting set of rules is minimized is analogous to the join-order optimization problem on relational database queries, which is well studied with many heuristic algorithms~\cite{DBLP:conf/sigmod/SelingerACLP79,DBLP:journals/vldb/SteinbrunnMK97}. However, most heuristics assume that the sizes of predicates are known in advance, since they consider sizes of only EDB predicates. We propose a new heuristic algorithm that is deterministic and well-suited to Datalog applications.
Our algorithm is presented below, where for each substep, we give the rationale.

\begin{itemize}
    \item If there is any pair of hypotheses such that the variables of one hypothesis is a subset of the variables of the other, we select any one such pair. This ensures that the intermediate rule has no added asymptotic complexity.

    \item Otherwise, apply the following steps in order, each step applied to the set of pairs selected so far, starting with the set of all pairs, until a unique pair is selected:

    \begin{enumerate}

    \item[(a.i)] In a rule, a variable is called \defn{removable} for two hypotheses if the variable only appears in those two hypotheses and nowhere else.  We select all pairs of hypotheses with the maximum number of removable variables.

    \item[(a.ii)] If domain size information is available, we multiply the sizes of the domains of each removable variable for each pair of hypotheses, and select all pairs with the maximum product.  Steps (a.i) and (a.ii) help to minimize the matching of the different values of the removable variables with other hypotheses in the rest of the rule.

    \item[(b)] We select all pairs of hypotheses that contain the maximum number of shared variables between the hypotheses in the pair. This helps to minimize the number of facts iterated over during the evaluation of the intermediate rule to be created.



    \item[(c.i)] We select all pairs of hypotheses that contain the maximum number of EDB hypotheses. This helps to best understand the complexity of the intermediate rule because sizes of EDB predicates are input parameters.

    \item[(c.ii)] We select all pairs of hypotheses in which the product of the sizes of the EDB hypotheses in the pair is the minimum. This helps to minimize the cost of the intermediate rule.

    \item[(d)] As a last resort, we select the pair of the leftmost two hypotheses.

    \end{enumerate}
\end{itemize}
Next, we show applications of this algorithm coupled with evaluation and complexity calculation.

\subsection{Andersen's pointer analysis for object-oriented languages and its complexity}
\label{subsec:andersen}

Pointer analysis comes in many flavors depending on what it takes into account. An \emph{intraprocedural} analysis only considers a single procedure. An \emph{interprocedural} analysis considers multiple procedures and interactions among them. A \emph{flow-insensitive} analysis does not take control flows into account, whereas a \emph{flow-sensitive} analysis does and produces a may-point-to or must-point-to set for each variable at each program point. A \emph{context-sensitive} analysis takes calling contexts into account, and produces a may-point-to or must-point-to set for each variable for each possible calling context. A \emph{context-insensitive} analysis does not.

For large programs, it is generally understood that flow-sensitivity and context-sensitivity are not feasible. The most well-known flow- and context-insensitive pointer analysis was developed by Andersen~\cite{phd:andersen}, and it is considered to offer a sweet spot between precision and efficiency~\cite{DBLP:conf/pldi/HardekopfL07}. Andersen formulates a may-point-to analysis in terms of type theory, and the formulation corresponds directly to a logical specification in Datalog. This subsection considers an intraprocedural pointer analysis in Datalog for an object-oriented language based on Andersen's analysis as described in~\cite{DBLP:journals/ftpl/SmaragdakisB15}.  We show the decomposition of the rules and calculate the precise time complexity.

\myparag{Predicates and rules for Andersen's analysis for OO languages}

Each statement has a corresponding fact, shown below.  For the first statemement, \mycode{h}, called a \defn{heap abstraction}, is a new constant created as an abstraction for the set of possible heap objects created by \mycode{new} when executing the statement, and \mycode{m}, not used in this intraprocedural analysis but used in later analyses, is the method containing the statement:\\
{\small
\begin{tabular}{l@{\hspace{10ex}}l}
 \hline
 \mycode{v = new Obj()} & \mycode{alloc(v,h,m)}\\
 \mycode{v = v2} & \mycode{move(v,v2)}\\
 \mycode{v.f = v2} & \mycode{store(v,f,v2)}\\
 \mycode{v = v2.f} & \mycode{load(v,v2,f)}\\
 \hline
\end{tabular}}
For example, method \mycode{foo} in Section~\ref{sec:intro} has three facts: \mycode{alloc(o1,h1,foo)}, \mycode{alloc(o2,h2,foo)}, and \mycode{move(o2,o1)},
where \mycode{h1} and \mycode{h2} are fresh constants.

The analysis defines the following two predicates and infers facts of them using the Datalog rules \mycode{(R1)}--\mycode{(R4)} in Fig.~\ref{fig:ruleinit}; additional explanations can be found in~\cite{DBLP:journals/ftpl/SmaragdakisB15}.

\begin{itemize}
    \item \mycode{\vpt(v,h)}: variable \mycode{v} may point to heap abstraction \mycode{h}
    \item \mycode{\fpt(h1,f,h2)}: heap abstraction \mycode{h1} may have its field \mycode{f} pointing to heap abstraction \mycode{h2}
\end{itemize}

\begin{figure}
\figrule
\begin{code}
v_pt(v,h) \IF alloc(v,h,m).                               (R1)
v_pt(v,h) \IF move(v,v2), v_pt(v2,h).                     (R2)
v_pt(v,h) \IF load(v,v2,f), v_pt(v2,h2), f_pt(h2,f,h).    (R3)
f_pt(h,f,h2) \IF store(v,f,v2), v_pt(v,h), v_pt(v2,h2).   (R4)

int1(v,f,h2) \IF load(v,v2,f), v_pt(v2,h2).  (R3/1)  // two decomposed rules for (R3)
v_pt(v,h) \IF int1(v,f,h2), f_pt(h2,f,h).    (R3/2)  //
int2(f,v2,h) \IF store(v,f,v2), v_pt(v,h).   (R4/1)  // two decomposed rules for (R4)
f_pt(h,f,h2) \IF int2(f,v2,h), v_pt(v2,h2).  (R4/2)  //
\end{code}
\figrule
\caption{An intraprocedural may-point-to analysis in Datalog, (R1)--(R4), and decomposed rules}
\label{fig:ruleinit}
\end{figure}

\myparag{Decomposition of rules and complexity analysis}

Using our algorithm for rule decomposition in Section~\ref{subsec:decompose}, we decompose rules \mycode{(R3)} and \mycode{(R4)} as shown in Fig.~\ref{fig:ruleinit}. For our rule decompositions, we note that in the input programs, there are more program points than variables, and more variables than heap abstractions. For \mycode{(R3)}, by algorithm step (a.ii), we select the first two hypotheses because \mycode{v2} is the removable variable whose domain size is maximum. For \mycode{(R4)}, we
can select the first two hypotheses or the first and the third because \mycode{v} and \mycode{v2} are removable and their domains are the same, so we select the leftmost two hypotheses that remove \mycode{v}.

We calculate precise complexities for the decomposed rules, and show the results in Table~\ref{tab:compinit}. For the rest of the paper, $p$ is the number of program points, $v$ is the number of variables, $h$ is the number of heap abstractions (i.e., \mycode{\#alloc}), and $f$ is the number of fields.

The sizes of IDB predicates are bounded by both the complexities of the rules inferring their facts, and the product of the sizes of the domains of their arguments. Sizes \mycode{\#int1} and \mycode{\#int2} are bounded by the complexities of \mycode{(R3/1)} and \mycode{(R4/1)} respectively,
because each firing produces at most one new fact;
they are also bounded by \sinO{v\times h\times f} based on the domains of their arguments. In all cases, \mycode{\#\vpt} is bounded by \sinO{v\times h}, and \mycode{\#\fpt} is bounded by \sinO{h^2\times f}. These complexities can be factored in when calculating the overall complexities.

Now, we give some insight into the calculated complexities. If we consider the worst case when all predicates are maximized (i.e., they have facts for all possible combinations of their argument values), the complexity of this analysis would be $O(p \times h^2)$. Under various conditions, we can obtain better complexities. For example, if all 
variables point to a constant number of heap abstractions, i.e., \sincsO{\vpt.2/1}=\sinO{1}, then the complexity would be \sinO{p\times h}.
If, in addition, all 
fields of variables point to a constant number of heap abstractions, i.e., \sincsO{\fpt.3/1,2}=\sinO{1}, then the complexity would be linear in the program size, \bigO{p}.

\begin{table}
\caption{Complexities for an intraprocedural may-point-to analysis}\label{tab:compinit}
\programmath
\begin{tabular}{ll}
\hline
\mycode{(R1)} &   \sincsO{alloc}  \\
\mycode{(R2)} &  \ofour{move}{\vpt.2/1}{\vpt}{move.1/2} \\
\mycode{(R3/1)} &  \ofour{load}{\vpt.2/1}{\vpt}{load.1,3/2} \\
\mycode{(R3/2)} & \ofour{int1}{\fpt.3/1,2}{\fpt}{int1.1/2,3}  \\
\mycode{(R4/1)} &  \ofour{store}{\vpt.2/1}{\vpt}{store.2,3/1} \\
\mycode{(R4/2)} & \ofour{int2}{\vpt.2/1}{\vpt}{int2.1,3/2}  \\
\hline
\end{tabular}
\unprogrammath
\end{table}

\section{Handling rules with function symbols---applied to context-sensitive may-point-to~analyses}
\label{sec:context}

A context-sensitive may-point-to analysis separates may-point-to sets for executions that map to different contexts, thereby increasing precision. One can consider different types of contexts such as call sites~\cite{phd:shivers,SharirPnueli:1981}, objects~\cite{DBLP:journals/tosem/MilanovaRR05}, and types~\cite{DBLP:conf/popl/SmaragdakisBL11}.
A rule-based model of context-sensitive analysis is presented in~\cite{DBLP:journals/ftpl/SmaragdakisB15}, but the rules contain function symbols, invalidating the polynomial-time evaluation guarantee for pure Datalog; different restrictions to the function symbols are provided to ensure polynomial-time evaluation.  In this section, we consider the restriction for the most sophisticated analysis, show how to extend our evaluation to handle function symbols, and calculate the complexities.

\myparag{Evaluation of Datalog with bounded-size terms}

We extend Datalog so that arguments of predicates may be terms, where a \defn{term} is either a constant, a variable, or a function symbol with arguments that are terms. We denote function symbols with uppercase letters\footnote{In logic programming, the converse is true; we use this notation to emphasize their presence.}, and require that each function symbol \mycode{F} be of fixed arity, so all occurrences of \mycode{F} take the same number of arguments. If a term is a function symbol \mycode{F} with arguments, we call it a \defn{term of \mycode{F}}. The number of constants and function symbols in a term is called its \defn{size}. The introduction of function symbols to Datalog rules makes the language Turing-complete, therefore invalidating complexity and termination guarantees~\cite{DBLP:journals/jlp/SchreyeD94}. We introduce a sufficient condition for detection of termination in the presence of function symbols, and discuss the evaluation and complexity when termination is guaranteed.

We say that a rule is \defn{size-bounding} for \mycode{F} if the sizes of the terms of \mycode{F} in the conclusion are guaranteed to be no larger than the size of the term of \mycode{F} with the maximum size in the hypotheses. If a set of Datalog rules extended with function symbols is size-bounding for every function symbol, then bottom-up evaluation is guaranteed to terminate. Note that rules with no function symbols in the conclusion are size-bounding by definition.

Given a set of size-bounding Datalog rules, we perform bottom-up evaluation and calculate its complexity exactly as described before. However, the sizes of predicates and domains of predicate arguments need to be made more precise for calculating the number of firings, because they can take on terms as values. For a function symbol \mycode{F}, we define \mycode{count(F)} to be the number of different terms of \mycode{F} that can appear during evaluation. If a size-bounding rule $r$ has a term of \mycode{F} in the conclusion distinct from terms of \c{F} that appear in its hypotheses, then the contribution of $r$ to \mycode{count(F)} is bounded by the product of the domains of the variables that appear in the term of \mycode{F} in the conclusion of $r$. Therefore, \mycode{count(F)} is bounded by the sum of such contributions in every rule. The size of the domain of the $i$th argument of a predicate \mycode{p} (\mycode{\#dom(p.i)}) is bounded by \mycode{count(F)}, if there is a rule whose conclusion's predicate is \mycode{p} and (i) its $i$th argument is a term of \mycode{F},
or (ii) there is a hypothesis of predicate \mycode{q} whose $j$th argument \mycode{a$_j$} is bounded by \mycode{count(F)} and the $i$th argument of the conclusion is \mycode{a$_j$}.

\myparag{2-call-site sensitive analysis with a 1-call-site sensitive heap}

A pointer analysis is said to be $n$-call-site sensitive if it tracks the last $n$ method calls leading to the execution of a statement, with an $m$-call-site sensitive heap if it tracks the last $m$ method calls leading to the creation of a heap object. Out of three context-sensitive analyses in~\cite{DBLP:journals/ftpl/SmaragdakisB15}, we consider the most complex one, a 2-call-site sensitive analysis with a 1-call-site sensitive heap.  The following
additional kinds of facts are used:\\
{\small
\begin{tabular}{ll}
\hline
 \mycode{vcall(v,s,p,m)} & virtual call \mycode{v.s(...)} is at program point \mycode{p} in method \mycode{m} \\
 \mycode{htype(h,t)}     & heap abstraction \mycode{h} has type \mycode{t} \\
 \mycode{lookup(t,s,m)}  & method with signature \mycode{s} of type \mycode{t} is \mycode{m} \\
 \mycode{this(m,t)}      & \mycode{this} variable for method \mycode{m} is \mycode{t} \\
 \mycode{farg(m,n,a)}    & method \mycode{m}'s \mycode{n}th formal argument is \mycode{a} \\
 \mycode{aarg(p,n,a)}    & program point \mycode{p} is a call whose \mycode{n}th actual argument is \mycode{a} \\
 \mycode{fret(m,v)}      & method \mycode{m}'s formal return variable is \mycode{v} \\
 \mycode{aret(p,v)}      & program point \mycode{p} is a call that assigns to actual return variable \mycode{v} \\
 \mycode{astore(v1,v2)}  & store into array element as in \mycode{v1[..] = v2} \\
 \mycode{aload(v1,v2)}   & load from array element as in \mycode{v1 = v2[..]} \\
 \mycode{etype(t,et)}    & array type \mycode{t} has element type \mycode{et} \\
 \mycode{stype(t1,t2)}   & type \mycode{t1} is a subtype of type \mycode{t2} \\
 \mycode{throw(p,v)}     & program point \mycode{p} throws variable \mycode{v} \\
 \mycode{catch(t,p,v)}   & for exceptions at program point \mycode{p} with arg type \mycode{t}, assign arg to \mycode{v} \\ 
 \mycode{in(p,m)}         & program point \mycode{p} is in method \mycode{m} \\
\hline
\end{tabular}
}

The analysis defines the following predicates and infers facts of them using the rules in Fig.~\ref{fig:rulecontext}. 
\begin{itemize}
    \item \mycode{\vpt(v,c,h,hc)}: variable \mycode{v} in context \mycode{c} may point to heap abstraction \mycode{h} in heap context \mycode{hc}$\!$ 
    \item \mycode{\fpt(h1,hc1,f,h2,hc2)}: heap abstraction \mycode{h1} in heap context \mycode{hc1} may have its field \mycode{f} pointing to heap abstraction \mycode{h2} in heap context \mycode{hc2}
    \item \mycode{r(m,c)}: method \mycode{m} is reached in context \mycode{c}
    \item \mycode{call(p,c1,m,c2)}: program point \mycode{p} in context \mycode{c1} calls method \mycode{m} in context \mycode{c2}
    \item \mycode{assign(v1,c1,v2,c2)}: variable \mycode{v1} in context \mycode{c1} is assigned the value of \mycode{v2} in context \mycode{c2}
\end{itemize}

\noindent Each rule gives a direct implication based on the meaning of the predicate. For example, rule \mycode{(R15)} says: if method \mycode{m} is reached in context \mycode{P(a,b)}, and variable \mycode{v} is assigned a new heap abstraction \mycode{h} in method \mycode{m}, then \mycode{v} in context \mycode{P(a,b)} may point to \mycode{h} in heap context \mycode{a}.
Note that a context is a pair represented with function symbol \mycode{P} since the analysis is 2-call-site sensitive, and that an initial fact \mycode{r(main,P(null,null))} can be used to indicate that method \mycode{main} is reached in an initial context where the last two calls before calling \mycode{main} are \mycode{null}.

Following the method above, we first show that the rules are size-bounding for \mycode{P} (the only function symbol). \mycode{(R16), (R17), (R18), (R20), (R21)} are trivially size-bounding since they have no function symbols in the conclusion. \mycode{(R15)} is size-bounding since the term of \mycode{P} appearing in the conclusion is identical to the one in the hypotheses. \mycode{(R19)} is size-bounding since (i) \mycode{P(a,b)} in the conclusion is identical to an occurrence of \mycode{P} in the hypotheses, and (ii) for \mycode{P(p,a)}, \mycode{p} is a program point (i.e., a constant), and therefore cannot have a larger size 
than \mycode{b}, and the size of 
\mycode{P(p,a)} is no more than \mycode{P(a,b)}.

Next, we determine which arguments of which predicates are bounded by \mycode{count(P)}. These are \mycode{\vpt.2} due to \mycode{(R15)} and \mycode{(R19)}, \mycode{r.2} due to \mycode{(R19)}, \mycode{call.2} and \mycode{call.4} due to \mycode{(R19)}, and \mycode{assign.2} and \mycode{assign.4} due to \mycode{(R20)} and  \mycode{(R21)} .

Finally, we determine \mycode{count(P)}. The only rule whose conclusion contains a term of \mycode{P} distinct from the terms in its hypotheses is \mycode{(R19)}. The variables of this term of \mycode{P} are \mycode{p} and \mycode{a}. The source of \mycode{p} is the third argument of the first hypothesis, therefore its domain size is \mycode{\#vcall.3}. The source of \mycode{a} is the first argument in the terms of \mycode{P} in the second and third hypotheses, but this argument of \mycode{P} is, as just analyzed, only from the third argument of the first hypothesis. Therefore, \mycode{count(P)} is $O((\mycode{\#vcall.3})^2)$.

We decompose the rules with our algorithm as shown in Fig.~\ref{fig:rulecontext}, and the calculated complexities are shown in Table~\ref{tab:compcontext}.

\begin{figure}
\figrule
\begin{code}
v_pt(v,P(a,b),h,a) \IF r(m,P(a,b)), alloc(v,h,m).                                 (R15)
v_pt(v,c,h,hc) \IF move(v,v2), v_pt(v2,c,h,hc).                                   (R16)
f_pt(h1,hc1,f,h2,hc2) \IF store(v1,f,v2), v_pt(v2,c,h2,hc2), v_pt(v1,c,h1,hc1).   (R17)
v_pt(v,c,h,hc) \IF load(v,v2,f), v_pt(v2,c,h2,hc2), f_pt(h2,hc2,f,h,hc).          (R18)
r(m,P(p,a)), v_pt(t,P(p,a),h,hc), call(p,P(a,b),m2,P(p,a)) \IF vcall(v,s,p,m1),
     r(m1,P(a,b)), v_pt(v,P(a,b),h,hc), htype(h,ht), lookup(ht,s,m2), this(m2,t).(R19)
assign(v1,c2,v2,c1) \IF call(p,c1,m,c2), farg(m,n,v1), aarg(p,n,v2).              (R20)
assign(v1,c1,v2,c2) \IF call(p,c1,m,c2), aret(p,v1), fret(m,v2).                  (R21)
v_pt(v,c,h,hc) \IF assign(v,c,v2,c2), v_pt(v2,c2,h,hc).                           (R22)

// decomposed rules:
int23(v1,v2,h1,hc1,h2,hc2) \IF v_pt(v2,c,h2,hc2), v_pt(v1,c,h1,hc1).      (R17/1)
f_pt(h1,hc1,f,h2,hc2) \IF store(v1,f,v2), int23(v1,v2,h1,hc1,h2,hc2).     (R17/2)
int24(v,f,c,h2,hc2) \IF load(v,v2,f), v_pt(v2,c,h2,hc2).                  (R18/1)
v_pt(v,c,h,hc) \IF int24(v,f,c,h2,hc2), f_pt(h2,hc2,f,h,hc).              (R18/2)
int25(s,p,m1,P(a,b),h,hc) \IF vcall(v,s,p,m1), v_pt(v,P(a,b),h,hc).       (R19/1)
int26(s,p,P(a,b),h,hc) \IF int25(s,p,m1,P(a,b),h,hc), r(m1,P(a,b)).       (R19/2)
int27(s,p,P(a,b),h,hc,ht,m2) \IF int26(s,p,P(a,b),h,hc), lookup(ht,s,m2). (R19/3)
int28(s,p,P(a,b),h,hc,m2) \IF int27(s,p,P(a,b),h,hc,ht,m2), htype(h,ht).  (R19/4)
r(m,P(p,a)), v_pt(t,P(p,a),h,hc), call(p,P(a,b),m2,P(p,a)) \IF
                            int28(s,p,P(a,b),h,hc,m2), this(m2,t).       (R19/5)
int29(c1,m,c2,n,v2) \IF call(p,c1,m,c2), aarg(p,n,v2).                    (R20/1)
assign(v1,c2,v2,c1) \IF int29(c1,m,c2,n,v2), farg(m,n,v1).                (R20/2)
int30(c1,m,c2,v1) \IF call(p,c1,m,c2), aret(p,v1).                        (R21/1)
assign(v1,c1,v2,c2) \IF int30(c1,m,c2,v1), fret(m,v2).                    (R21/2)
\end{code}
\figrule
\caption{A 2-call-site sensitive analysis with a 1-call-site sensitive heap, and decomposed rules}\label{fig:rulecontext}
\end{figure}

\begin{small}
\begin{table}
\caption{Complexities for a 2-call-site sensitive analysis with a 1-call-site sensitive heap}\label{tab:compcontext}
\programmath
\begin{tabular}{ll}
\hline
\mycode{(R15)} & \ofour{r}{alloc.1,2/3}{alloc}{r.2/1}    \\
\mycode{(R16)} & \ofour{move}{\vpt.2,3,4/1}{\vpt}{move.1/2} \\
\mycode{(R17/1)} & \doubcsO{\vpt}{\vpt.1,3,4/2} \\
\mycode{(R17/2)} & \ofour{store}{int23.3,4,5,6/1,2}{int23}{store.2/1,3} \\
\mycode{(R18/1)} & \ofour{load}{\vpt.2,3,4/1}{\vpt}{load.1,3/2} \\
\mycode{(R18/2)} & \ofour{int24}{\fpt.4,5/1,2,3}{\fpt}{int24.1,3/2,4,5} \\
\mycode{(R19/1)} & \ofour{vcall}{\vpt.2,3,4/1}{\vpt}{vcall.2,3,4/1} \\
\mycode{(R19/2)} & \sincsO{int25} \\
\mycode{(R19/3)} & \ofour{int26}{lookup.1,3/2}{lookup}{int26.2,3,4,5/1} \\
\mycode{(R19/4)} & \sincsO{int27} \\
\mycode{(R19/5)} & \ofour{int28}{this.2/1}{this}{int28.1,2,3,4,5/6} \\
\mycode{(R20/1)} & \ofour{call}{aarg.2,3/1}{aarg}{call.2,3,4/1} \\
\mycode{(R20/2)} & \ofour{int29}{farg.3/1,2}{farg}{int29.1,3,5/2,4} \\
\mycode{(R21/1)} & \ofour{call}{aret.2/1}{aret}{call.2,3,4/1} \\
\mycode{(R21/2)} & \ofour{int30}{fret.2/1}{fret}{int30.1,3,4/2} \\
\hline
\end{tabular}
\unprogrammath
\end{table}
\end{small}

\section{Handling rules with universal quantification---applied to flow-sensitive must-point-to~analysis}
\label{sec:flow}

\renewcommand\c[1]{\texttt{\small #1}} 
\newcommand\mpt{must\_pt}
\newcommand\fmpt{f\_must\_pt}
\newcommand\cmpt{\mycode{must\_pt}}
\newcommand\cfmpt{\mycode{f\_must\_pt}}

Must-point-to analysis determines the heap abstractions that a pointer variable or expression must refer to, as opposed to may refer to, in all program executions.
Flow-sensitive analysis determines analysis results specific to each program point, as opposed to one global result for the program.
Therefore flow-sensitive must-point-to analysis can give significantly more certain results that complement flow-insensitive may-point-to alias analysis.
This analysis poses two new challenges:
\begin{enumerate}

\item The analysis is much more complex, requiring extensions to Datalog with universal quantification and negation.

\item The analysis algorithm is much more sophisticated, requiring new techniques to keep the complexity from increasing.
\end{enumerate}

\myparag{Specification using Datalog rules with universal quantification
  and negation}

The analysis is specified using seven
rules~\cite{DBLP:journals/ftpl/SmaragdakisB15}, shown in Fig.~\ref{fig:rulemust} (after the changes noted in the third paragraph below).  The last
two rules are the core of the analysis.  The first five rules
define \cmpt~and a simple case of \cfmpt, where \mycode{alloc}, \mycode{move}, \mycode{load}, and \mycode{store} are as in Section~\ref{subsec:andersen} except with an additional first argument indicating the program point, and \mycode{phi} is an instruction for merging the values of two variables.  The first five rules are simple
Datalog rules with one, two, or three hypotheses; they can be analyzed using
the method in Sections~\ref{subsec:complexity} and~\ref{subsec:decompose}, yielding a time complexity of
\sinO{p\times h^2}.
This section focuses on the two core rules,
which are Datalog extended with universal quantification, simple negation, as
well as inequality.  These two rules are the core of the flow-sensitive analysis because they infer field-must-point-to information for each program point by combining information from all its predecessor points.
\begin{figure}
\figrule
\begin{code}
\mpt(var,h) \IF alloc(_,var,h,_).
\mpt(to,h) \IF move(_,to,from), \mpt(from,h).
\mpt(to,h) \IF phi(_,to,from1,from2), \mpt(from1,h), \mpt(from2,h).
\mpt(to,h2) \IF load(i,to,v,f), \mpt(v,h), \fmpt(i,h,f,h2).
\fmpt(i,h,f,h2) \IF store(i,v,f,from), \mpt(from,h2), \mpt(v,h).
\fmpt(j,h,f,h2) \IF
     next(_,j), \fmpt(_,h,f,h2), (forall i: next(i,j) \THEN \fmpt(i,h,f,h2)),
     not store(j,_,f,_), not vcall(_,_,j,_), not alloc(j,_,h), not alloc(j,_,h2).
\fmpt(j,h,f,h2) \IF
     next(_,j), \fmpt(_,h,f,h2), (forall i: next(i,j) \THEN \fmpt(i,h,f,h2)),
     store(j,v,f,_), \mpt(v,h3), h3 != h.
\end{code}
\figrule
\caption{A flow-sensitive must-point-to analysis in Datalog with universal quantification, negation, and inequality}\label{fig:rulemust}
\end{figure}



The first core rule says that, just after instruction \mycode{j}, \mycode{h} must point
via its field \mycode{f} to \mycode{h2} if (1) \mycode{j} is the next instruction of
some instruction, (2) \mycode{h} must point via \mycode{f} to \mycode{h2} just after
some instruction, (3) for all instructions \mycode{i} just before \mycode{j},
\mycode{h} must point via \mycode{f} to \mycode{h2} at \mycode{i}, and (4) \mycode{j} is not a
\mycode{store}, \mycode{vcall}, or \mycode{alloc} instruction that can change the
must-point-to information.

The second core rule concludes the same if the same conditions hold except that
\mycode{j} is a \mycode{store} instruction into field \mycode{f} of \mycode{v}, and
\mycode{v} must point to a heap abstraction \mycode{h3} that is not \mycode{h}.

Note that, compared to the original two core rules~\cite{DBLP:journals/ftpl/SmaragdakisB15}, we added the first two hypotheses in each rule and moved the conditions about \mycode{j} out of the universal quantification.
The two new hypotheses bind the free variables not bound by the universal quantifications, and are necessary for the rules to be correct;
without them, the universal quantification returns
true when no \mycode{i} satisfies its domain condition \mycode{next(i,j)},
which would lead to \cfmpt\ to hold
for all values of \mycode{h}, \mycode{f}, and \mycode{h2} for any \mycode{j}
for which no \mycode{i} satisfies \mycode{next(i,j)}.
The conditions about \mycode{j} are moved out because
they do not depend on the universally quantified variable \mycode{i}.
These also show that the
analysis is complex and universal quantification is challenging.

\myparag{Analysis algorithm for universal quantification 
  and inequality} 

Despite negation and inequality in the core rules, the set of \cfmpt\
facts that can be inferred still increases monotonically.  Therefore,
the set can be computed as a least fixed point
as for Datalog.
However, if computed straightforwardly, universal quantification adds
a linear factor after each \cfmpt\ fact is added.  We show how to
compute it, as well as the negation and inequality, incrementally in
\bigO{1} time.

\notes{
  all x (p(x) -> r(y))
= all x (not p(x) or r(y))
= all x (not p(x)) or r(y)
= all x (p(x) -> true) or r(y)

x = i, y = j
  all x (p(x,y) -> q(x) and r(y))
= all x (not p(x,y) or (q(x) and r(y)))
= all x ((not p(x,y) or q(x)) and not p(x,y) or r(y)))
= all x ((p(x,y) -> q(x)) and (p(x,y) -> r(y)))
= all x (p(x,y) -> q(x)) and all x (p(x,y) -> r(y))
!= all x (p(x,y) -> q(x)) and r(y)

but we think it is equal if some x p(x,y)

an example:
all x (hardworking(x) -> goodgrade(x) and red(jon))
all x (hardworking(x) -> goodgrade(x)) and red(jon)
} 

%
%
%
Consider the universal quantification, in both core rules:
\begin{code}
   (forall i: next(i,j) \THEN \fmpt(i,h,f,h2))
\end{code}
To compute it efficiently, we maintain the following four auxiliary
invariants:
\begin{code}
   prev[j] = \{i: (i,j) in next\}  and
   prev_count[j] = \#prev[j],  for j in next.2

   prev_pt[j,h,f,h2] = \{i: (i,j) in next, (i,h,f,h2) in \fmpt\}  and
   prev_pt_count[j,h,f,h2] = \#prev_pt[j,h,f,h2],  for j in next.2, (h,f,h2) in \fmpt.2,3,4
\end{code}
and replace the universal quantification with the following \bigO{1}
time test between two aggregate count values:
\begin{code}
   prev_count[j] = prev_pt_count[j,h,f,h2]
\end{code}

Variables \mycode{prev} and \mycode{prev\_count} for the first two invariants are initialized by iterating over each
element \mycode{(i,j)} of input \mycode{next}, adding \mycode{i} to \mycode{prev[j]} and
incrementing \mycode{prev\_count[j]}, in a total of \bigO{\mycode{\#next}} time.
The next two invariants are maintained incrementally at addition of
\mycode{(i,h,f,h2)} to \cfmpt\, as follows, taking a total of
\bigO{\mycode{\#next.2/1}\times\cfmpt} time overall all additions:
\begin{code}
   for j in next[i]:                 // use next to get each next node
      prev_pt[j,h,f,h2] \m{\cup}= \{i\}    \m{\!}   // i is new to h,f,h2 because (i,h,f,h2) is new
      prev_pt_count[j,h,f,h2] += 1   // increment the corresponding count by 1
\end{code}

The first core rule now becomes Datalog with simple negations as \bigO{1}
time tests, and with the universal quantification as an \bigO{1} time
equality test between two counts.  Its total time complexity is
\bigO{\mycode{\#next.2}\times\mycode{\#\fmpt.2,3,4}}.

The second core rule is similar in terms of the universal quantification,
but it does not have simple negations but an inequality.  We handle
the inequality specially, replacing the last two hypotheses on the
last line with the following, removing the extra variable \mycode{h3}:
\begin{code}
   \mpt[v] - \{h\} != \{\}   // use \mpt[v] to get heap abstractions that v must point to
\end{code}
There is only one value for \mycode{v} in a \mycode{store} instruction,
and the above element subtraction and test take \bigO{1} time.  So the
total time complexity of this rule is again
\bigO{\mycode{\#next.2}\times\mycode{\#\fmpt.2,3,4}}.


\myparag{Complexity guarantees}

Summing all time complexities together, from initialization,
maintaining auxiliary invariants, and using the two resulting rules,
yields the total time complexity
\begin{code}
   \bigO{\mycode{#next} + \mycode{#next.2/1}\times\mycode{#\fmpt} + \mycode{#next.2}\times\mycode{#\fmpt.2,3,4}}
\end{code}
\mycode{\#next} is bounded by \bigO{p}, the size of the program.  \mycode{\fmpt.2,3,4}
is bounded by the domain sizes of its three arguments \bigO{h\times f\times h}.
Therefore, the total time complexity is \bigO{p\times h^2\times f}.

\section{Additional pointer analyses and summary of complexity analysis results}
\label{sec:summary}

Besides the 3 pointer analyses discussed, we also studied the remaining 6 analyses in~\cite{DBLP:journals/ftpl/SmaragdakisB15}, including 3 in Appendix A\tplpadd{
of~\cite{2016arXiv160801594T}};
we do not present the rest in detail because they are simpler and do not illustrate additional logic rule features for complexity analysis.
There are also other analyses that can be specified using Datalog rules, such as the context-free-language formulation in~\cite{DBLP:conf/popl/ZhengR08}. We believe that the reader can follow our method to produce a set of rules and analyze their complexities easily.

In addition to the precise complexities that we calculated, here we also present the worst-case complexities in simpler terms, and provide conditions under which the complexities are linear or quadratic.  Table~\ref{tab:summary} summarizes, for each analysis, the features used, maximum number of hypotheses in the rules, worst-case complexities, and complexities conditioned on constraints on sizes of predicates.
We denote an $n$-call-site sensitive analysis with an $m$-call-site sensitive heap, as $(n,m)$-context.
The conditions used are as follows, where the conditions on all the EDB predicates are typical for real programs.
\begin{description}
    \item (C1): \sincsO{\vpt.2/1}=\sinO{1}
    \item (C2): \sincsO{\fpt.3/1,2}=\sinO{1}
    \item (C3): \sincsO{lookup.1,3/2}=\sincsO{this.2/1}=\sincsO{call.2/1}=\sincsO{farg.3/1,2}=\sincsO{fret.2/1}=\sinO{1}
    \item (C4): \sincsO{htype.2/1}=\sincsO{in.2/1}=\sincsO{\tpt.2/1}=\sincsO{throw.2/1}=\sinO{1}
    \item (C5): \sincsO{\mpt.2/1}=\sinO{1}
    \item (C6): \sincsO{\fmpt.2,3,4}=\sinO{h}
\end{description}

\noindent (C1) says that each variable may point to a constant number of heap abstractions. (C2) says that each field of each variable may point to a constant number of heap abstractions.  The rest are similar.


Complexities of pointer analysis with constraints on the sizes of program parameters have been studied. One can obtain the complexities achieved in such studies using our method, and substituting the relevant complexity parameters in our analysis with the constraints. In~\cite{DBLP:conf/sas/SridharanF09}, the authors present an algorithm for Andersen's analysis, which runs in $O((v+h)^2)$ time for $k$-sparse programs, where $v$ is the number of variables and $h$ is the number of heap abstractions. The definition of $k$-sparse programs has two constraints: For our rules in Figure~\ref{fig:ruleinit}, the first constraint implies $\sincsO{store.2,3/1}=\sincsO{load.1,3/2}=O(1)$, and the second constraint implies that \mycode{\#move}+\mycode{\#int1}+\mycode{\#int2} $\le O(v+h)$. Substituting these constraints in our complexity analysis in Table~\ref{tab:compinit}, we obtain $O((v+h)\times (\mycode{\#\vpt.2/1}+\mycode{\#\fpt.3/1,2}))$, which is in the worst case $O((v+h)\times h)$. Thus, we obtain a better and more precise complexity than~\cite{DBLP:conf/sas/SridharanF09}.

For Andersen OO analysis, interprocedural analysis, and (0,1)-context-sensitive analysis, the worst-case complexities when parametrized by only program size, $n$, are known---an upper bound of $O(n^3)$ for the first two~\cite{phd:andersen} and $O(n^5)$ for the third~\cite{DBLP:conf/pldi/WilsonL95}. However, to our knowledge, we present more precise complexities for these analyses for the first time, and for the other analyses, we present complexities for the first time. Our complexity results are improvements since they are tighter than known worst-case complexities, and when our fine-grained analyses are used, the running time of the analyses can be better understood.

\afterpage{%
\bgroup
\def\arraystretch{1.2}
\begin{table}
\caption{Summary of complexities.
  Recall $p$, $h$, and $f$ are number of program points, heap abstractions, and fields, respectively.}\label{tab:summary}
\programmath
\begin{tabular}{@{}l@{}p{2.5cm}@{$\!\!$}c@{}l@{}l@{}}
\hline
Analysis & Features & \#Hypo. & Worst-case & Conditional \\
\hline
Andersen OO (Sec.~\ref{sec:base})  & Pure Datalog & 3 & \sinO{p\times h^2} & \sinO{p\times h} if (C1), \sinO{p} if (C1) and (C2)\\
Interprocedural (Sec.~A.1)\tablefootnote{\label{note1} In Appendix A\tplpadd{ of~\cite{2016arXiv160801594T}}.} & Pure Datalog & 6 & \sinO{p^2\times h} &  \sinO{p} if (C1), (C2) and (C3) \\
Arrays (Sec.~A.2)\footnoteref{note1} & Pure Datalog & 7 & \sinO{p^2\times h} & \sinO{p} if (C1), (C2) and (C3) \\
Exceptions (Sec.~A.3)\footnoteref{note1} & Negation & 5 & \sinO{p^2\times h}  & \sinO{p} if (C1) and (C3) \\
Reflection & Pure Datalog & 9 & \sinO{p^3\times h} & \sinO{p\times h} obtainable\tablefootnote{The conditions to obtain this complexity involve EDB predicates not discussed in this paper.}\\
(0,1)-context & Function symbols & 6 & \sinO{p^3\times h^2}  & -- \\
(1,1)-context & Function symbols & 6 & \sinO{p^4\times h^2} & -- \\
(2,1)-context (Sec.~\ref{sec:context}) & Function symbols & 6 & \sinO{p^5\times h^2} &  -- \\
Flow must (Sec.~\ref{sec:flow}) & Univ. quant., negation, inequality & 8 & \sinO{p\times h^2\times f} & \sinO{p\times h} if (C5) and (C6) \\
\hline
\end{tabular}
\unprogrammath
\end{table}
\egroup
}

\section{Related work and conclusion}

We discuss related work on Datalog evaluation, applications of rules to pointer analysis, precise complexities for pointer analyses, and directions for future work.

Evaluation of Datalog has been studied for a long time~\cite{Ceri:1990:LPD:83229}. Optimal bottom-up evaluation of Datalog rules with complexity guarantees is first given in~\cite{Liu:2009:DRE:1552309.1552311}, but no algorithm is given for decomposing rules except for trying all decompositions.  We build on this method for evaluating Datalog rules and calculating complexities, but extend it to handle rules with many hypotheses and other Datalog extensions. Our new algorithm and method are able to obtain new or more precise complexities compared with the best previous complexities, as discussed in Section~\ref{sec:summary}.

Formulation of various static program analyses as rules has been studied. In particular, Andersen's pointer analysis~\cite{phd:andersen} was formulated as deductive rules in~\cite{DBLP:conf/pldi/HeintzeT01a}, and given as logic rules in~\cite{DBLP:conf/ppdp/SahaR05}. Andersen's analysis with many flavors was given in a recent survey~\cite{DBLP:journals/ftpl/SmaragdakisB15}, on which we base our study. The fact that the time complexity of Andersen's analysis is worst-case cubic has been known since the original introduction. \cite{DBLP:conf/sas/SridharanF09} notes that typical behavior is different than worst case and proves that under certain conditions the analysis is quadratic.
We give the precise time complexities for Andersen's analysis for an object-oriented language, and show also precise
conditions, additional to existing literature, under which the complexities are linear or quadratic directly as results of our complexity analyses.  We also obtain precise complexities for pointer analyses for additional language features, and provide methods for handling extensions to Datalog such as function symbols when such extensions are necessary to implement the analyses.

Must-point-to analyses are more complex, and rules modeling the analyses involve universal quantification and inequality beyond pure Datalog. \cite{DBLP:journals/toplas/HindBCC99} gives an \sinO{n^5} algorithm for a must-alias analysis (closely related to must-point-to analyses); using methods that are also employed in optimal bottom-up evaluation of Datalog, \cite{DBLP:journals/lisp/Goyal05} improves this complexity to $O(n^3)$. In this paper, we show how to handle must-point-to analysis expressed using rules extended with universal quantification and a special inequality, and provide precise complexity analyses for our efficient implementation. Transforming quantifications into aggregate queries such as counts has been used in other applications, e.g., distributed algorithms~\cite{Liu+12DistPL-OOPSLA}, but how to handle inequality in general is a subject for future study.

Future directions include analyzing and optimizing the space complexity of pointer analyses, especially to remove unnecessary intermediate predicates introduced for rules with many hypotheses, and optimization of demand-driven pointer analysis via queries, e.g., by using the methods of~\cite{DBLP:conf/ppdp/TekleL10} and~\cite{DBLP:conf/sigmod/TekleL11}.

\label{lastpage}

\bibliographystyle{ACM-Reference-Format-Journals}
\bibliography{main}

\tplpdel{\clearpage

\appendix

\section{Pointer analyses for advanced features and their complexities}

Our method for handling many hypotheses applies also to pointer analyses for other advanced language features.
We present different analyses~\cite{DBLP:journals/ftpl/SmaragdakisB15} built on the base analysis in Fig.~\ref{fig:ruleinit}.
For each analysis, we apply our algorithm for rule decomposition as described before, including for rules with simple negation, and show the complexity calculation.

A more precise pointer analysis can be obtained by making the analysis interprocedural. The rules for an interprocedural version of Andersen's analysis are more complex, and contain up to 6 hypotheses.  We present the rule decompositions and calculated complexities for this analysis in Section~\ref{appsec:inter}.

Another feature to add is to analyze arrays.  The analysis includes using a rule with 7 hypotheses.  We present the rule decompositions and calculated complexities for analyzing arrays in Secction~\ref{appsec:array}.

We also show an analysis of exceptions. This analysis uses a simple form of negation where only some EDB hypotheses are negated. Such negation is handled during evaluation by simply testing the truth value for the negated hypotheses. We describe the handling of negation during evaluation and rule decomposition, and then present the rule decompositions and calculated complexities for analyzing exceptions in Section~\ref{appsec:exception}.

\subsection{An interprocedural may-point-to analysis}
\label{appsec:inter}

In addition to the IDB predicates for the base analysis presented in Section~\ref{subsec:andersen}, the following IDB predicates are used for interprocedural analysis.
\begin{itemize}
    \item \mycode{r(m)}: Method \mycode{m} is reached.
    \item \mycode{call(p,m)}: Statement at program point \mycode{p} calls method \mycode{m}.
    \item \mycode{assign(v1,v2)}: \mycode{v1} is assigned \mycode{v2} either as part of argument or return passing.
\end{itemize}

These IDB predicates are defined with the rules in Fig.~\ref{fig:inter}. In addition, \mycode{(R1)} is redefined with \mycode{(R1*)} to be more precise, and more precise definitions for \mycode{\vpt} are added. Since \mycode{(R5)}, \mycode{(R6)} and \mycode{(R7)} have more than two hypotheses, we decompose them using our algorithm resulting in the rules also shown in Fig.~\ref{fig:inter}, and we calculate the complexities are shown in Table~\ref{tab:inter}.

\begin{figure}
\figrule
\begin{code}
v_pt(v,h) \IF  alloc(v,h,m), r(m).                                        (R1*)
r(m2), v_pt(t,h), call(p,m2) \IF vcall(v,s,p,m1), r(m1), v_pt(v,h),
                                   htype(h,ht), lookup(ht,s,m2), this(m2,t). (R5)
assign(v1,v2) \IF call(p,m), farg(m,n,v1), aarg(p,n,v2).                      (R6)
assign(v1,v2) \IF call(p,m), aret(p,v1), fret(m,v2).                          (R7)
v_pt(v1,h) \IF assign(v1,v2), v_pt(v2,h).                                     (R8)

int3(v,s,p) \IF vcall(v,s,p,m1), r(m1).                        (R5/1)
int4(s,p,h) \IF int3(v,s,p), v_pt(v,h).                        (R5/2)
int5(p,h,ht,m2) \IF int4(s,p,h), lookup(ht,s,m2).              (R5/3)
int6(p,h,m2) \IF int5(p,h,ht,m2), htype(h,ht).                 (R5/4)
r(m2), v_pt(t,h), call(p,m2) \IF int6(p,h,m2), this(m2,t).     (R5/5)
int7(m,n,v2) \IF call(p,m), aarg(p,n,v2).                      (R6/1)
assign(v1,v2) \IF int7(m,n,v2), farg(m,n,v1).                  (R6/2)
int8(m,v1) \IF call(p,m), aret(p,v1).                          (R7/1)
assign(v1,v2) \IF int8(m,v1), fret(m,v2).                      (R7/2)
\end{code}
\figrule
\caption{Additional and redefined rules for an interprocedural analysis, and decomposed rules}\label{fig:inter}
\end{figure}

\begin{table}
\caption{Complexities for an interprocedural point-to analysis}\label{tab:inter}
\programmath
\begin{tabular}{ll}
\hline
Rule                & Complexity\\
\hline
\mycode{(R1*)} &   \sincsO{alloc}  \\
\mycode{(R5/1)} &  \sincsO{vcall}  \\
\mycode{(R5/2)} &  \ofour{int3}{\vpt.2/1}{\vpt}{int3.2,3/1}  \\
\mycode{(R5/3)} &  \ofour{int4}{lookup.1,3/2}{lookup}{int4.2,3/1}  \\
\mycode{(R5/4)} &   \sincsO{int5}   \\
\mycode{(R5/5)} &  \ofour{int6}{this.2/1}{this}{int6.1,2/3}  \\
\mycode{(R6/1)} &  \ofour{call}{aarg.2,3/1}{aarg}{call.2/1}  \\
\mycode{(R6/2)} &  \ofour{int7}{farg.3/1,2}{farg}{int7.3/1,2}  \\
\mycode{(R7/1)} &  \ofour{call}{aret.2/1}{aret}{call.2/1}  \\
\mycode{(R7/2)} &  \ofour{int8}{fret.2/1}{fret}{int8.2/1}  \\
\mycode{(R8)} &  \ofour{assign}{\vpt.2/1}{\vpt}{assign.1/2}  \\
\hline
\end{tabular}
\unprogrammath
\end{table}

\subsection{May-point-to analysis for arrays}
\label{appsec:array}

Analysis of arrays is done in an \defn{array-insensitive} manner, so that the analysis is performed on arrays but different array indices are not distinguished. A new IDB predicate \mycode{\apt(a,h)} is defined, indicating that elements of array \mycode{a} may point to heap abstraction \mycode{h}, and a new rule is added for \mycode{\vpt} as shown in Fig.~\ref{fig:array}. In the figure, we show the decomposition of these rules using our algorithm. We calculate the complexities from these rules as shown in Table~\ref{tab:array}.

\begin{figure}
\figrule
\begin{code}
a_pt(h1,h2) \IF astore(v1,v2), v_pt(v1,h1), v_pt(v2,h2), htype(h1,ht1),
               htype(h2,ht2), etype(ht2,et), stype(ht1,et).            (R9)
v_pt(v,h) \IF aload(v,v2), v_pt(v2,h2), a_pt(h2,h).                     (R10)

int9(v2,h1) \IF astore(v1,v2), v_pt(v1,h1).      (R9/1)
int10(h1,h2) \IF int9(v2,h1), v_pt(v2,h2).       (R9/2)
int11(ht1,ht2) \IF etype(ht2,et), stype(ht1,et). (R9/3)
int12(h1,ht2) \IF htype(h1,ht1), int11(ht1,ht2). (R9/4)
int13(h1,h2) \IF int12(h1,ht2), htype(h2,ht2).   (R9/5)
a_pt(h1,h2) \IF int10(h1,h2), int13(h1,h2).      (R9/6)
int14(v,h2) \IF aload(v,v2), v_pt(v2,h2).        (R10/1)
v_pt(v,h) \IF int14(v,h2), a_pt(h2,h).           (R10/2)
\end{code}
\figrule
\caption{Additional rules for analyzing arrays, and decomposed rules}\label{fig:array}
\end{figure}

\begin{table}
\caption{Complexities incurred by analyzing arrays}\label{tab:array}
\programmath
\begin{tabular}{ll}
\hline
Rule                & Complexity\\
\hline
\mycode{(R9/1)} & \ofour{astore}{\vpt.2/1}{\vpt}{astore.2/1} \\
\mycode{(R9/2)} & \ofour{int9}{\vpt.2/1}{\vpt}{int9.2/1} \\
\mycode{(R9/3)} & \ofour{etype}{stype.1/2}{stype}{etype.1/2} \\
\mycode{(R9/4)} & \ofour{htype}{int11.2/1}{int11}{htype.1/2} \\
\mycode{(R9/5)} & \ofour{int12}{htype.1/2}{htype}{int12.1/2} \\
\mycode{(R9/6)} & \sincsO{int10} \\
\mycode{(R10/1)} & \ofour{aload}{\vpt.2/1}{\vpt}{aload.1/2} \\
\mycode{(R10/2)} & \ofour{int14}{\apt.2/1}{\apt}{int14.1/2} \\
\hline
\end{tabular}
\unprogrammath
\end{table}

\subsection{Handling negation---applied to for may-point-to analysis for exceptions}
\label{appsec:exception}


A hypothesis can be \defn{negated}, indicated by a preceding \mycode{not}. Given a rule with negated hypotheses, and values for the variables, the rule infers the conclusion as a fact if all non-negated hypotheses are true and all negated hypotheses are false for the given values of variables. We say that rules have \defn{simple negation} if the only negated hypotheses in the rules are EDB hypotheses; this negation is simple because it can be performed as a check after all non-negated hypotheses are processed. Therefore, a rule only needs decomposition if there are more than two non-negated hypotheses.

In the case when a negated hypothesis appears in a rule that needs to be decomposed, we apply our algorithm as described before without considering the negated hypotheses, and if at any point all the essential variables of a negated hypotheses appear in an intermediate rule, then we add the negated hypotheses to the rule. A variable of a hypothesis is \defn{essential} in a rule if it appears in any other hypothesis or conclusion; we name non-essential variables with an underscore (\mycode{\_}) in the rules. After this decomposition, during bottom-up evaluation of a rule with negated hypotheses, for each value combination of variables of non-negated hypotheses that make the hypotheses true, we check if the negated hypotheses hold for the given values of the essential variables and any value of the non-essential variables. If none of the negated hypotheses hold, then we add the conclusion as a fact. This does not change the complexity of the rule since such checks can be performed in constant time.

\myparag{Exception analysis} Rules are added to allow analysis of the exception flow of a program as shown in Fig.~\ref{fig:except}. The rules define a new IDB predicate \mycode{\tpt(m,h)} capturing what heap abstractions \mycode{h} a method \mycode{m} can throw to its catchers for an exception.  Rules are also added \mycode{\vpt} for analyzing the exceptions. The rules contain simple negation since \mycode{catch} is an EDB predicate. We show the decomposition of the rules as well in Fig.~\ref{fig:except}, and the complexities calculated with our method in Table~\ref{tab:except}.

\begin{figure}
\figrule
\begin{code}
t_pt(m,h) \IF in(p,m), throw(p,v), v_pt(v,h), htype(h,ht), not catch(ht,p,_).  (R11)
t_pt(m,h) \IF in(p,m), call(p,m2), t_pt(m2,h), htype(h,ht), not catch(ht,p,_). (R12)
v_pt(v2,h) \IF throw(p,v), v_pt(v,h), htype(h,ht), catch(ht,p,v2).             (R13)
v_pt(v,h) \IF call(p,m), t_pt(m,h), htype(h,ht), catch(ht,p,v).                (R14)

int15(p,h) \IF throw(p,v), v_pt(v,h).                      (R11/1)
int16(p,h) \IF int15(p,h), htype(h,ht), not catch(ht,p,_). (R11/2)
t_pt(m,h) \IF in(p,m), int16(p,h).                         (R11/3)
int17(p,h) \IF call(p,m2), t_pt(m2,h).                     (R12/1)
int18(p,h) \IF int17(p,h), htype(h,ht), not catch(ht,p,_). (R12/2)
t_pt(m,h) \IF in(p,m), int18(p,h).                         (R12/3)
int19(v,ht,v2) \IF throw(p,v), catch(ht,p,v2).             (R13/1)
int20(h,ht,v2) \IF v_pt(v,h), int19(v,ht,v2).              (R13/2)
v_pt(v2,h) \IF int20(h,ht,v2), htype(h,ht).                (R13/3)
int21(m,ht,v) \IF call(p,m), catch(ht,p,v).                (R14/1)
int22(h,ht,v) \IF int21(m,ht,v), t_pt(m,h).                (R14/2)
v_pt(v,h) \IF int22(h,ht,v), htype(h,ht).                  (R14/3)
\end{code}
\figrule
\caption{Aadditional rules for analyzing exceptions, and decomposed rules}\label{fig:except}
\end{figure}

\begin{table}
\caption{Complexities incurred by analyzing exceptions}\label{tab:except}
\programmath
\begin{tabular}{ll}
\hline
Rule        & Complexity\\
\hline
\mycode{(R11/1)} & \ofour{throw}{\vpt.2/1}{\vpt}{throw.1/2} \\
\mycode{(R11/2)} & \ofour{int15}{htype.2/1}{htype}{int15.1/2} \\
\mycode{(R11/3)} & \ofour{in}{int16.2/1}{int16}{in.2/1} \\
\mycode{(R12/1)} & \ofour{call}{\tpt.2/1}{\tpt}{call.2/1} \\
\mycode{(R12/2)} & \ofour{int17}{htype.2/1}{htype}{int17.1/2} \\
\mycode{(R12/3)} & \ofour{in}{int18.2/1}{int18}{in.2/1} \\
\mycode{(R13/1)} & \ofour{throw}{catch.1,3/2}{catch}{throw.2/1} \\
\mycode{(R13/2)} & \ofour{\vpt}{int19.2,3/1}{int19}{\vpt.2/1} \\
\mycode{(R13/3)} & \sincsO{int20} \\
\mycode{(R14/1)} & \ofour{call}{catch.1,3/2}{catch}{call.2/1} \\
\mycode{(R14/2)} & \ofour{int21}{\tpt.2/1}{\tpt}{int21.2,3/1} \\
\mycode{(R14/3)} & \sincsO{int22} \\

\hline
\end{tabular}
\unprogrammath
\end{table}
}

\end{document}